\documentclass[12pt]{iopart}
\usepackage{graphicx,iopams}
\usepackage{hyperref}
 
\begin{document}

\title[]{Conductance distributions in chaotic mesoscopic cavities}

\author{Santosh Kumar and Akhilesh Pandey}

\address{School of Physical Sciences, Jawaharlal Nehru University, 
New Delhi - 110067, India}
\eads{\mailto{skumar.physics@gmail.com}, \mailto{ap0700@mail.jnu.ac.in}}

\begin{abstract}
We consider the conductance distributions in chaotic mesoscopic cavities for all three invariant classes of random matrices for arbitrary number of channels $N_1, N_2$ in the connecting leads. We show that the Laplace transforms of the distributions can be expressed in terms of determinants in the unitary case and Pfaffians in the orthogonal and symplectic cases. The inverse Laplace transforms then give the exact distributions. This formalism is particularly useful for small values of $N=\min(N_1,N_2)$, and thus is of direct experimental relevance. We also obtain the conductance distributions for orthogonal-unitary and symplectic-unitary crossover ensembles.
\end{abstract}


\maketitle

\section{Introduction}

The study of quantum transport properties in chaotic mesoscopic cavities has attracted a great deal of attention in the last two decades \cite{MK}-\cite{YA}. The importance of theoretical investigation in this direction has increased in recent years because of the availability of sophisticated experimental techniques which can be used to test these theoretical predictions \cite{YA,CBP,CCM,HPM,HHZ,BLM,SSML}. Landauer-B\"uttiker formalism provides a powerful way to investigate the quantum transport properties in these systems \cite{Lndr,FL,Btkr}. This formalism is based on the scattering matrix approach and enables one to work out important quantities such as conductance and shot-noise power from the knowledge of the transmission eigenvalues. These transmission eigenvalues, in turn, are obtained from the polar-decomposition of the scattering matrix \cite{MK,Bnkr,BM,JPB}.

Quantum dots are examples of mesoscopic cavities where one can investigate the electronic transport properties experimentally and compare them with the theoretical results \cite{YA,CBP,CCM,HPM}. Experiments on these systems now involve the study of conductance and shot-noise moments as well as their full distributions. Apart from the quantum dots, microwave cavities serve as important systems where the theoretical predictions can be compared with experimental results. In microwave experiments one studies the transmission of electric field through the cavity \cite{HHZ,BLM,SSML}. The same theory applies to both the cases because of the equivalence of mathematical structures of the (time-independent) Schr\"odinger and Helmholtz equations. Moreover, the experiments on microwave cavities are free from the complicating effects of thermal fluctuations etc. which lead to deviations from the standard fully coherent theory in quantum dots. 

These chaotic mesoscopic cavities belong to the class of complex systems where the particularity of the microscopic details are rendered irrelevant by the complexity of the system and the macroscopic behaviour is decided solely by the associated global symmetries. Random matrix theory (RMT) has been successfully applied to the study of such complex systems. In the case of chaotic mesoscopic cavities the scattering matrix is modelled using a unitary matrix belonging to the circular ensembles of random matrices. The appropriate random matrix ensemble is decided by the time-reversal and spin-rotational symmetry properties of the cavity \cite{MK,Bnkr}. The classification of random matrix ensembles follows from their invariance under orthogonal, unitary and symplectic transformations and are accordingly referred to as orthogonal ensemble (OE), unitary ensemble (UE) and symplectic ensemble (SE) \cite{RM}. OE and SE are applicable to time-reversal invariant systems which are with and without rotational symmetry respectively. UE, on the other hand, applies to systems where time-reversal is not a good symmetry. These ensembles are also designated by the Dyson index $\beta$ which assumes the values 1, 2 and 4 for OE, UE and SE respectively.

It has been shown that under the RMT treatment the statistics of transmission eigenvalues is identical to that of a special case of Jacobi random matrix ensembles \cite{Bnkr,AM,For,SKP}. This identification leads to simplifications in calculating explicitly the above mentioned physical quantities. Exact results are available for the averages and variances of conductance and shot-noise power for arbitrary number of channels (or modes) $N_1, N_2$ in the leads connected to the cavity \cite{MK,Bnkr,SKP,SS,Nov1,Nov2}. However, significant progress in deriving the exact distributions of these quantities for arbitrary $N_1, N_2$ has been made only recently. For the conductance distribution the explicit expressions have been given for $N_1=N_2=1,2$ in orthogonal ($\beta=1$) and unitary ($\beta=2$) cases \cite{MK,Bnkr,BM,JPB}. The solution for arbitrary $N_1, N_2$ for $\beta=2$ was also described in terms of Toda lattice and Painlev\'e equations \cite{OK1}. The asymptotic case of large $N_1=N_2$ has been analysed in \cite{OK1,VMB}. Further progress in this direction has been made by calculating higher cumulants of conductance and shot-noise power \cite{Nov1,Nov2,SWS,SSW,OK2}. Recently Khoruzhenko, Savin and Sommers have studied in detail the conductance and shot-noise distributions for all $\beta$ and arbitrary $N_1, N_2$ \cite{KSS}. They have obtained the results in terms of a Fourier series with terms involving Pfaffians. However, explicit forms for the conductance distribution are given only for $N=1,2$ with arbitrary $M$, where $N=\min(N_1,N_2)$ and $M=\max(N_1,N_2)$.

It is known that the distribution takes the form of a Gaussian for large $N$ values \cite{Pol}. For small $N$, however, significant departures from the Gaussian behaviour are observed. This is important from the point of view of experimental studies on chaotic cavities \cite{CBP,CCM,HPM,HHZ,BLM,SSML}.  It is therefore desirable to have explicit results for small $N$ values.  Our primary aim in this paper is to provide explicit results for an important range of small $N$ values. Using results in \cite{RM}, we show that the conductance distribution can be given for all $N_1, N_2$ in terms of inverse Laplace transform of a determinant for UE and Pfaffians in OE and SE cases. This formalism is particularly useful for finding the explicit expressions of the distributions for all three $\beta$ for values of $N$, say upto 6. We also obtain the Laplace transform results for OE-UE and SE-UE crossover ensembles.

The paper is organized as follows. In section 2 we use Landauer formula and joint probability density (JPD) of transmission eigenvalues to obtain the Laplace transform of conductance distribution for all $\beta$ and $N_1, N_2$ values. In section 3 we give explicit results for a range of small $N$ values by taking inverse Laplace transform of the results in section 2. Section 4 deals with the conductance distribution in OE-UE and SE-UE crossover ensembles. We conclude in section 5 with the summary of our results and some general remarks.

\section{Laplace Transform of Conductance Distribution}
\label{secLT}

The JPD of transmission eigenvalues $\{T_j\}$ is known from random matrix theory as \cite{MK,Bnkr,For,SKP}
\begin{equation}
\label{Pbeta}
P^{(\beta)}(T_1,...,T_N)=C_N^{(\beta)} |\Delta_N(T_1,...,T_N)|^\beta \prod_{j=1}^N T_j^\alpha,
\end{equation}
where $0\leq T_j \leq 1$, $\Delta_N(T_1,...,T_N)=\prod_{j<k}(T_j-T_k)$ is the Vandermonde determinant and $\alpha=\beta(|N_1-N_2|+1)/2-1$. The normalization $C^{(\beta)}_N$ is found using Selberg's integral \cite{RM} as
\begin{equation}
C_N^{(\beta)}=\prod_{j=0}^{N-1}\frac{\Gamma(1+\frac{\beta}{2})\Gamma(2+\alpha+\frac{\beta}{2}(N+j-1))}{\Gamma(1+\frac{\beta}{2}(1+j))\Gamma(1+\alpha+\frac{\beta}{2}j)\Gamma(1+\frac{\beta}{2}j)}.
\end{equation}
The dimensionless conductance at zero temperature is given in terms of the transmission eigenvalues using the Landauer formula as \cite{Bnkr,Lndr,FL}
\begin{equation}
g=\sum_{j=1}^N T_j.
\end{equation}
Thus the problem of finding the conductance distribution reduces to the mathematical problem of performing the following integral:
\begin{equation}
\label{F-beta}
F^{(\beta)}(g)=\int_0^1\!\!\cdots\!\int_0^1 \!\!\!\delta\Big(g-\sum_{j=1}^N T_j\Big)P^{(\beta)}(T_1,...,T_N)dT_1...dT_N.
\end{equation}
One natural way to deal with the delta function in the integrand is to take the Laplace transform of (\ref{F-beta}). We find
\begin{equation}
\label{LF-beta}
\widetilde{F}^{(\beta)}(s)=\int_0^1\!\!\cdots\!\int_0^1 \!\!\!C_N^{(\beta)}|\Delta_N|^\beta \prod_{j=1}^N e^{-s T_j}T_j^\alpha dT_1...dT_N.
\end{equation}
Here $\widetilde{F}^{(\beta)}(s)=\mathcal{L}\{F^{(\beta)}(g)\}$ is the Laplace transform of $F^{(\beta)}(g)$. Such averages involving $|\Delta_N|^\beta$ have already been considered in chapter 5 of \cite{RM}. We work along similar lines to evaluate the above integral.

For $\beta=1$ the main complication in solving (\ref{LF-beta}) comes from the absolute value sign of $\Delta_N$. We resolve this difficulty by using the method of integration over alternate variables \cite{RM}. We find 
\begin{equation}
\label{LF1e}
\widetilde{F}^{(1)}(s)=\Gamma(N+1)C_N^{(1)}\mbox{Pf}[\Psi_{j,k}^{(1)}(s)]_{j,k=0,...,N-1},
\end{equation}
when $N$ is even, and
\begin{equation}
\label{LF1o}
\widetilde{F}^{(1)}(s)=\Gamma(N+1)C_N^{(1)}\mbox{Pf}
            \left[\begin{array}{cc}
             \Psi_{j,k}^{(1)}(s) & \Phi_j^{(1)}(s) \\
            -\Phi_k^{(1)}(s)  &  0
            \end{array}\right]_{j,k=0,...,N-1},        
\end{equation} 
when $N$ is odd. Here Pf$[\mathcal{A}]$ is the Pfaffian of the even-dimensional antisymmetric matrix $\mathcal{A}$ \cite{RM}. In (\ref{LF1e}) and (\ref{LF1o}) $\Psi_{j,k}^{(1)}$ and $\Phi_j^{(1)}$ are given by
\begin{equation}
\label{Psi1}
\Psi_{j,k}^{(1)}(s)=\int_{0}^1\int_{0}^1 \mbox{sgn}(x-y)\,e^{-sx}e^{-sy}x^{\alpha+j}y^{\alpha+k}\,dx\,dy,
\end{equation}
\begin{equation}
\Phi_j^{(1)}(s)=\int_0^1 e^{-sx}x^{\alpha+j} dx.
\end{equation}
Note that $\Psi_{j,k}^{(1)}(s)=-\Psi_{k,j}^{(1)}(s)$. We carry out the integration in (\ref{Psi1}) by considering the ranges $0\leq y \leq x$ and $x\leq y \leq 1$, thereby dealing with the $\mbox{sgn}(x-y)$ factor.

The $\beta=2$ case is comparatively easier to handle and the result is obtained in terms of a determinant. We get
\begin{equation}
\label{LF2}
\widetilde{F}^{(2)}(s)=\Gamma(N+1)C_N^{(2)}\mbox{Det}[\Psi_{j,k}^{(2)}(s)]_{j,k=0,...,N-1},
\end{equation}
where $\Psi_{j,k}^{(2)}(s)$ is given by
\begin{equation}
\label{Psi2}
\Psi_{j,k}^{(2)}(s)=\int_{0}^1 e^{-sx}x^{\alpha+j+k}\,dx.
\end{equation}
The $\Psi_{j,k}$ in this case is symmetric between the indices $j,k$. 

For $\beta=4$ we again obtain the result in terms of a Pfaffian, viz.,
\begin{equation}
\label{LF4}
\widetilde{F}^{(4)}(s)=\Gamma(N+1)C_N^{(4)}\mbox{Pf}[\Psi_{j,k}^{(4)}(s)]_{j,k=0,...,2N-1}.
\end{equation}
Here $\Psi_{j,k}^{(4)}(s)$ is given by
\begin{equation}
\label{Psi4}
\Psi_{j,k}^{(4)}(s)=\int_{0}^1 e^{-sx}x^{\alpha+j+k-1}(k-j)\,dx.
\end{equation}
$\Psi_{j,k}$ is again antisymmetric under the exchange of $j,k$ as in $\beta=1$ case.

The conductance distributions for the respective cases follow from the inverse Laplace transforms of (\ref{LF1e}), (\ref{LF1o}), (\ref{LF2}) and (\ref{LF4}). Explicit results are possible because of the specific forms of the integrals that appear in the evaluation of $\Psi$ and $\Phi$. As mentioned earlier these results are well suited for finding the distribution of conductance for small values of $N$ (upto 6 or 7) and one can use symbolic manipulation software package like Mathematica. The corresponding values of $N_1$ and $N_2$ cover almost all the combinations of number of channels that are typically considered in experiments \cite{CBP,CCM,HPM,HHZ}. 

We remark that one can find the moments of conductance from the expansion of Laplace transform:
\begin{equation}
\widetilde{F}(s)=\sum_{\mu=0}^\infty \frac{(-1)^\mu}{\Gamma(\mu+1)}s^\mu \left<g^\mu\right>.
\end{equation}
Here $\left<g^\mu\right>$ represents the $\mu$th moment of conductance, given by
\begin{equation}
 \left<g^\mu\right>=\int_0^\infty g^\mu F(g)\,dg.
\end{equation}
Similarly the cumulants can be found using the expansion of $\log(\widetilde{F}(s))$.

\section{Exact Results for Conductance Distribution}

For $N=1, 2$ with arbitrary $M$, the exact results for all the three $\beta$ cases are already known \cite{KSS}. We have for $N=1$ and arbitrary $M$,
\begin{equation}
F^{(\beta)}(g)=\frac{\beta M}{2}\, g^{\beta M/2-1},
\end{equation}
for $0\le g \le 1$ and zero otherwise. For $N=2$ and arbitrary $M$ we have
\begin{eqnarray}
\label{N2M}
\fl
\nonumber
F^{(\beta)}(g)=M g^{\beta M-1}\Big[\frac{\Gamma(\beta(M+1)/2+1)\,\Gamma(\beta M/2)}{\Gamma(\beta/2)\,\Gamma(\beta M)}-(-1)^{(\beta M -1)/2}\frac{\Gamma(\beta(M+1)/2+1)}{\Gamma(\beta)\,\Gamma(\beta (M-1)/2)}\\
\times\Theta(g-1)\sum_{j=0}^{\beta}{\beta \choose j} B_{1-g}\left(j+\beta(M-1)/2, 1-\beta M\right)\Big],
\end{eqnarray}
for $0\le g \le 2$ and zero otherwise. Here $\Theta(z)$ is the Heaviside step function and $B_z(a,b)$ is the incomplete beta-function. Further simplification can be made from (\ref{N2M}) for $\beta=1$ and one obtains
\begin{equation}
F^{(1)}(g)=\frac{1}{2}M(M+1)\Big[\Big(\frac{g}{2}\Big)^{M-1}-(g-1)^{(M-1)/2}\Theta(g-1)\Big].
\end{equation}
For other $\beta$ values one has to evaluate (\ref{N2M}) for each $M$ separately.

We give below explicit results for the conductance distribution for small $N_1,N_2$ values. Note that $F^{(\beta)}(g)=0$ for all $g<0$ and $g>N$. We have considered $N\ge2$ and $M\leq 5$ for $\beta=1,2$ and $N\ge2$ and $M\leq 4$ for $\beta=4$. It is possible to work out explicit results for larger values of $N$ also. However, the results become progressively lengthier. We have also included $N=2$ results below as the general result (\ref{N2M}) gives compact expressions for low values of $M$. 

\subsection{Results for OE (\texorpdfstring{$\beta=1$}{beta=1})}
For $\beta=1$, we obtain the following results:
\begin{itemize}
\item
$M=2,N=2$
\begin{equation}
F^{(1)}(g)=\frac{3}{2}[g-2(g-1)^{1/2}\Theta(g-1)], ~~~~~~~~~~~~~~~~~ 0\leq g \leq 2 ~.
\end{equation}

\item
$M=3,N=2$
\begin{equation}
F^{(1)}(g)=\frac{3}{2}[g^2-4(g-1)\Theta(g-1)], ~~~~~~~~~~~~~~~~~~~ 0\leq g \leq 2 ~, \\
\end{equation}

\item
$M=3,N=3$
\begin{equation}
\fl
F^{(1)}(g)=\cases{
     \frac{6}{7}g^{7/2}, & ~~~~~~~~~~~~~~~~~~~~~~~$0\leq g \leq 1$, \\
     \frac{3}{28} [35 g^3-175 g^2+273 g-125\\
     -8(g-2)^{5/2}(g+5)\Theta(g-2)], & ~~~~~~~~~~~~~~~~~~~~~~~$1\leq g \leq 3$.     
     }
\end{equation}

\item
$M=4,N=2$
\begin{equation}
\fl
F^{(1)}(g)=\frac{5}{4}[g^3-8(g-1)^{3/2}\Theta(g-1)], ~~~~~~~~~~~~~~~~~~~~~~~~~~~~~~~~~~~ 0\leq g \leq 2 , \\
\end{equation}

\item
$M=4,N=3$
\begin{eqnarray}
\fl
\nonumber
F^{(1)}(g)=\frac{3}{8}[g^5-(g-1)^3(g^2-12g+51)\Theta(g-1)\\
-(g-2)^3(g^2+6g+24)\Theta(g-2)], ~~~~~~~~~~~~~~~~~~~~~~ 0\leq g \leq 3,
\end{eqnarray}

\item
$M=4,N=4$
\begin{equation}
\fl
F^{(1)}(g)=\cases{
\frac{5}{27456}[429g^7-512(g-1)^{9/2}\\
\times(6g^2-64g+201)\Theta(g-1)],& $0\leq g \leq 2$ ,\\
-\frac{5}{27456}[429g^7-72072g^5+672672g^4\\
-2800512g^3+6150144g^2-6935552g\\
+3158016-1024(g-3)^{11/2}(3g+4)\Theta(g-3)],& $2\leq g \leq 4$.
}
\end{equation}

\item
$M=5,N=2$
\begin{equation}
F^{(1)}(g)=\frac{15}{16}[g^4-16(g-1)^2\Theta(g-1)], ~~~~~~~~~~~~~~~~ 0\leq g \leq 2, \\
\end{equation}

\item
$M=5,N=3$
\begin{equation}
\fl
F^{(1)}(g)=\cases{
     \frac{20}{143}g^{13/2}, & $0\leq g \leq 1$ , \\
     \frac{5}{2288} [3003 g^5-21021 g^4+55770 g^3\\
     -70070g^2+42315g-9933\\
     -32(g-2)^{7/2}(2g^3+14g^2+63g+231)\Theta(g-2)], & $1\leq g \leq 3$ .     
     }
\end{equation}

\item
$M=5,N=4$
\begin{equation}
\fl
F^{(1)}(g)=\cases{
     \frac{5}{384}[g^9-12(g-1)^6 (3g^2-30g+83)\Theta(g-1)\\
     -2(g-2)^5 (g^4-8g^3+168g^2-608g\\
     +2608)\Theta(g-2)], & ~~~~$0\leq g \leq 3$,\\
     -\frac{5}{384}(g-4)^9, &~~~ $3\leq g \leq 4$.     
     }
\end{equation}
\end{itemize}

\subsection{Results for UE (\texorpdfstring{$\beta=2$}{beta=2})}

For $\beta=2$ we get the conductance distributions as:
\begin{itemize}

\item
$M=2,N=2$
\begin{equation}
\fl
F^{(2)}(g)=\cases{
     2g^3, & ~~~~~~~~~~~~~~~~~~~~~~~~~~~~~~~~~~~~~~~~~~~~~~~~$0\leq g \leq 1$, \\
     -2(g-2)^3, & ~~~~~~~~~~~~~~~~~~~~~~~~~~~~~~~~~~~~~~~~~~~~~~~~$1\leq g \leq 2$.
     }
\end{equation}

\item
$M=3,N=2$
\begin{equation}
\fl
F^{(2)}(g)=\cases{
     \frac{6}{5}g^5, & ~~~~~~~~~~~~~~~~~~~~~~~~~~~~~~~$0\leq g \leq 1$, \\
     -\frac{6}{5} (g-2)^3 (g^2+6g-6), & ~~~~~~~~~~~~~~~~~~~~~~~~~~~~~~~$1\leq g \leq 2$.
     }
\end{equation}

\item
$M=3,N=3$
\fl
\begin{equation}
\fl
F^{(2)}(g)=\cases{
     \frac{3}{14} [g^8-3(g-1)^4 (g^4-4 g^3+62 g^2\\
     -228 g+309)\Theta(g-1)], & ~~~~~~~~~~~~~~~~~~~$0\leq g \leq 2$,\\     
     \frac{3}{14} (g-3)^8, & ~~~~~~~~~~~~~~~~~~~$2\leq g \leq 3$.
     }
\end{equation}

\item
$M=4,N=2$
\begin{equation}
\fl
F^{(2)}(g)=\cases{
     \frac{4}{7}g^7, &  ~~~~~~~~~~~~$0\leq g \leq 1$, \\
     -\frac{4}{7} (g-2)^3 (g^4+6g^3+24g^2-60g+30), & ~~~~~~~~~~~~$1\leq g \leq 2$.
     }
\end{equation}

\item
$M=4,N=3$
\begin{equation}
\fl
F^{(2)}(g)=\cases{
     \frac{2}{77} [g^{11}-3(g-1)^6 (g^5+6g^4-89g^3+936g^2\\
     -3174g+3860)\Theta(g-1)], & ~~~~~~~~$0\leq g \leq 2$,\\     
     \frac{2}{77} (g-3)^8 (g^3+24g^2-6g-60), & ~~~~~~~~$2\leq g \leq 3$.
     }
\end{equation}

\item
$M=4,N=4$
\begin{equation}
\fl
F^{(2)}(g)=\cases{     
     \frac{2}{3003}[g^{15}-4(g-1)^9(g^6-6g^5+330g^4-4010g^3\\
     +25110g^2-69516g+73116)\Theta(g-1) \\
     +2(g-2)^7(3 g^8- 48 g^7+ 1596 g^6- 16464 g^5\\
     + 89880 g^4 - 294336 g^3+ 970256 g^2 - 2196032 g\\
     +2758628)\Theta(g-2)], & ~~$0\leq g \leq 3$,\\
 -\frac{2}{3003}(g-4)^{15}& ~~$3\leq g \leq 4$.
     }
\end{equation}

\item
$M=5,N=2$
\begin{eqnarray}
\fl
\nonumber
F^{(2)}(g)=\frac{5}{21}[g^9-2(g-1)^4(g^5+4g^4+10g^3\\
+20g^2-280g+560)\Theta(g-1)],~~~~~~~~~~~~~~~~~~~~~~~~~~~~  0\leq g \leq 2.
\end{eqnarray}

\item
$M=5,N=3$
\begin{equation}
\fl
F^{(2)}(g)=\cases{
     \frac{5}{2002} [g^{14}-(g-1)^8 (3g^6 + 24g^5 + 108g^4\\
      - 3280g^3 + 31930g^2 - 104640g+120900)\Theta(g-1)], & $0\leq g \leq 2$,\\     
     \frac{5}{2002} (g-3)^8(g^6+ 24g^5 + 324g^4-400g^3\\
      - 570g^2 - 1920g + 3900), & $2\leq g \leq 3$.
     }
\end{equation}

\item
$M=5,N=4$
\begin{equation}
\fl
F^{(2)}(g)=\cases{
     \frac{5}{415701}[g^{19}-2(g-1)^{12}(2 g^7+24 g^6-699 g^5\\
 +19538 g^4-209310 g^3+1127760 g^2-2855515 g\\
 +2799990)\Theta(g-1)+6(g-2)^9 (g^{10}+18g^9-675g^8\\
 +15000g^7-116910g^6+342828g^5+779824g^4\\
 -10900080g^3+54224220g^2-135038120g\\
 +154989420)\Theta(g-2)], & $0\leq g \leq 3$, \\
 -\frac{5}{415701}(g-4)^{15}(g^4+ 60 g^3+ 210 g^2\\
 - 940 g +180), & $3\leq g \leq 4$.
 }
\end{equation}

\end{itemize}

\subsection{Results for SE (\texorpdfstring{$\beta=4$}{beta=4})}
For $\beta=4$ our results are as follows:

\begin{itemize}
\item
$M=2,N=2$
\begin{equation}
\fl
F^{(4)}(g)=\cases{
     \frac{12}{7}g^7, & ~~~~~~~~~~~~~~~~~~~~~~~~~$0\leq g \leq 1 $, \\
     -\frac{12}{7}(g-2)^5(g^2+10g-10), & ~~~~~~~~~~~~~~~~~~~~~~~~~$1\leq g \leq 2$.
     }
\end{equation}

\item
$M=3,N=2$
\begin{equation}
\fl
F^{(4)}(g)=\cases{
     \frac{4}{11}g^{11}, & ~~~~~~~~~~~~~$0\leq g \leq 1 $, \\
     -\frac{4}{11} (g-2)^5 (g^6+10g^5+60g^4+280g^3\\
-1190g^2+1260g-420), &~~~~~~~~~~~~ $1\leq g \leq 2 $.
     }
\end{equation}

\item
$M=3,N=3$
\begin{equation}
\fl
F^{(4)}(g)=\cases{
     \frac{15}{2431}[g^{17}- (g-1)^8(3 g^9+ 24 g^8 - 844 g^7\\
     + 17496 g^6 - 167990 g^5 + 1013400 g^4 \\
     - 3859020 g^3+ 8836968 g^2- 10975377 g \\
     +5747952 )\Theta(g-1)], & ~~~~~~~~~~~~$0\leq g \leq 2 $,\\
     \frac{15}{2431}(g-3)^{14} (g^3+ 42 g^2- 7 g-112), & ~~~~~~~~~~~~$2\leq g \leq 3 $.    
     }
\end{equation}

\item
$M=4,N=2$
\begin{eqnarray}
\fl
\nonumber
F^{(4)}(g)=
     \frac{8}{143}[g^{15} - 2 (g-1)^6 (g^9 + 6g^8 + 21g^7 + 56g^6 \\
     \nonumber
     + 126g^5+ 252g^4 - 29568g^3 + 155232g^2 \\
     - 288288g+192192)\Theta(g-1)], ~~~~~~~~~~~~~~~~~~~~~~~~~~~~ 0\leq g \leq 2.
\end{eqnarray}

\item
$M=4,N=3$
\begin{eqnarray}
\fl
\nonumber
F^{(4)}(g)=\frac{60}{1062347}[g^{23} - 3 (g-1)^{12} (g^{11}+ 12 g^{10}+ 78 g^9\\
\nonumber
+ 364 g^8- 73017 g^7 + 1534512 g^6- 14807352 g^5\\
\nonumber
+ 83363952 g^4- 287746326 g^3+ 597826824 g^2\\
\nonumber
 - 685194300 g +334840968) \Theta(g-1)\\
 \nonumber
     + 3 (g-2)^9 (g^{14}+ 18 g^{13}+ 180 g^{12}+ 1320 g^{11}\\
 \nonumber    
     - 140844 g^{10}+ 2208888 g^9- 15548456 g^8\\
  \nonumber   
     + 62418048 g^7- 143468556 g^6+ 209723976 g^5\\
  \nonumber   
     - 212061696 g^4+ 91459872 g^3+ 783610044 g^2\\
     - 2453722488 g+2560102776) \Theta(g-2)], ~~~~~~~~~~~~~~~~ 0\leq g \leq 3.
\end{eqnarray}
\end{itemize}

\begin{figure*}[ht]
\centering
\includegraphics*[width=0.92 \textwidth]{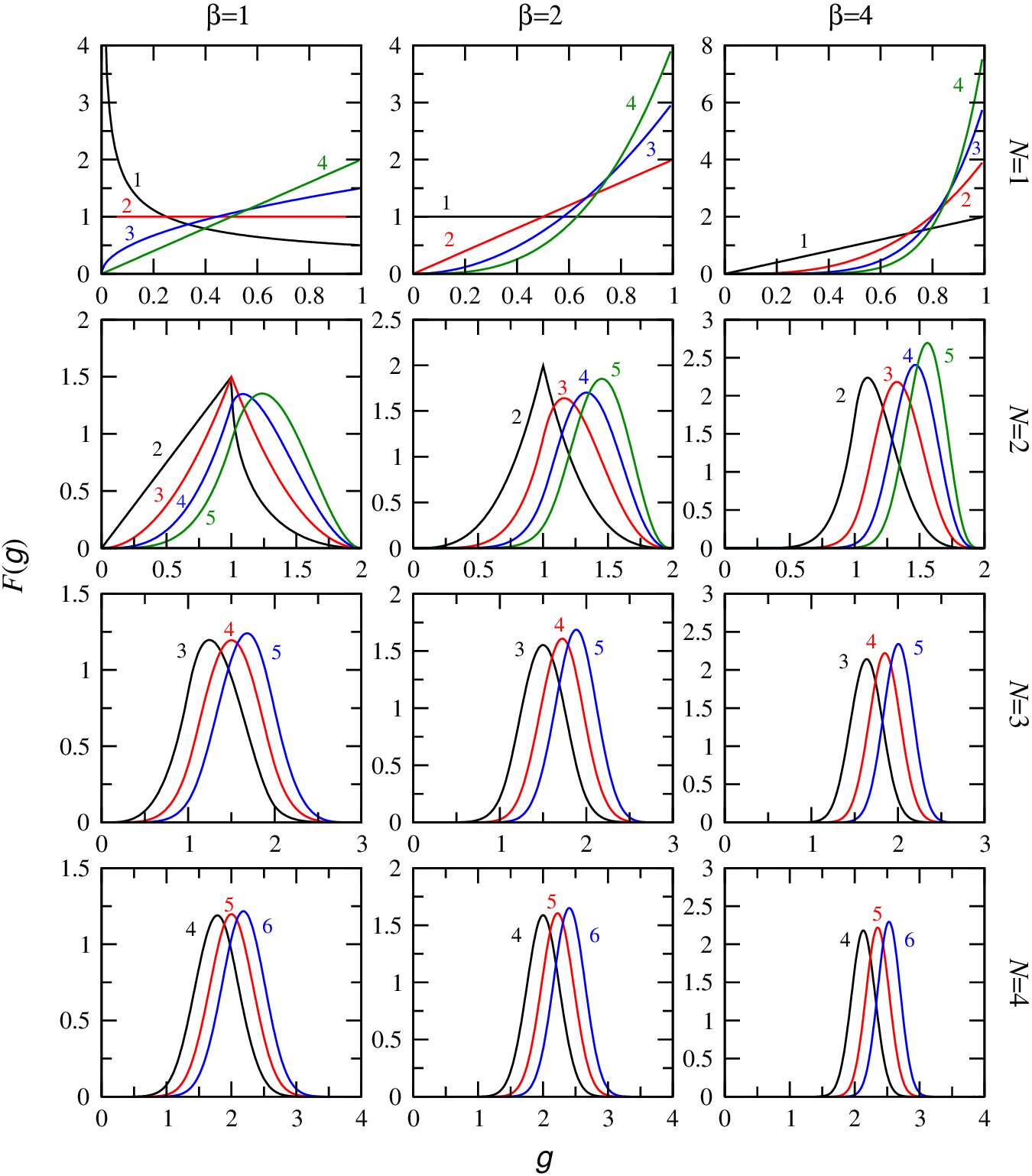}
\caption[]{Conductance distributions for $\beta=1, 2$ and 4 for various $M, N$. The three columns correspond to $\beta=1,2,4$ whereas the the rows correspond to increasing $N$. The numbers near the curves represent $M$ values.}
\end{figure*}

The plots of conductance distribution for these and other higher values of $N_1,N_2$ have been shown in figure 1. As mentioned above the conductance distribution for large $N$ can be approximated by the Gaussian distribution \cite{Pol}
\begin{equation}
\label{Gauss}
F_G(g)=\frac{1}{\sqrt{2\pi \mathrm{var}(g)}}e^{-(g-\left<g\right>)^2/2\mathrm{var}(g)}.
\end{equation}
Here $\left<g\right>$ and $\mbox{var}(g)$ are the average conductance and the variance of conductance respectively and are given by \cite{MK,Bnkr,SKP,KSS}
\begin{equation}
\left<g\right> =\frac{N_1 N_2}{(N_\mathrm{S}-1+2/\beta)},
\end{equation}
\begin{equation}
 \mathrm{var}(g)=\frac{2N_1 N_2(N_1-1+2/\beta)(N_2-1+2/\beta)}{\beta(N_\mathrm{S}-2+2/\beta)(N_\mathrm{S}-1+2/\beta)^2(N_\mathrm{S}-1+4/\beta)},
\end{equation}
 where $N_\mathrm{S}=N_1+N_2$.
For large $N_1,N_2$ these get reduced to
\begin{equation}
\left<g\right> =\frac{N_1 N_2}{N_\mathrm{S}}, 
\end{equation}
\begin{equation}
 \mathrm{var}(g)=\frac{2N_1^2 N_2^2}{\beta N_\mathrm{S}^4}.
\end{equation}
Comparison between the exact results and the above Gaussian approximation of (\ref{Gauss}) has been shown in figure 2 for $N\geq 2$ and $M\leq 4$. The departure is shown in figure 3 as the relative percentage difference $100(F(g)-F_G(g))/F(g)$. It has been shown recently that the Gaussian approximation holds in the range $N/4\leq g \leq 3N/4$ for large $N$ \cite{OK1,VMB}. This Gaussian range is also observed in figure 3, even though the $N$ values are rather small. Outside this range the Gaussian approximation becomes poor. We also find that the power law approximations \cite{VMB} suggested outside this range do not work well for these $N$ values.

\begin{figure*}[ht]
\centering
\includegraphics*[width=0.92 \textwidth]{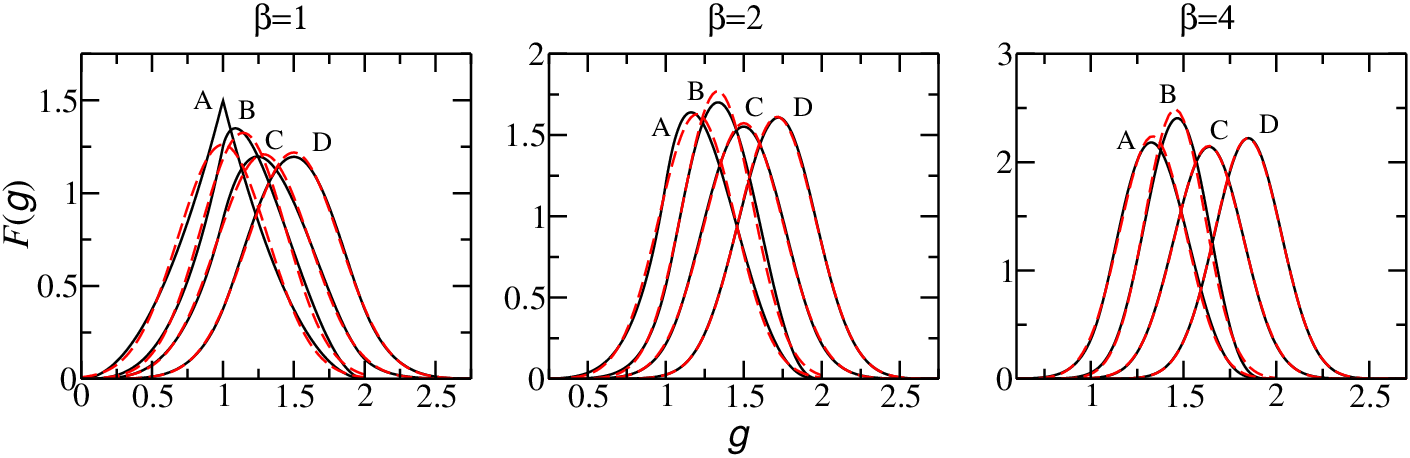}
\caption[]{Comparison between the exact results (solid lines, black) and Gaussian approximations (broken lines, red) given by equation (\ref{Gauss}). The labels near the curves represent the $M,N$ values as A: $M=3,N=2$, B: $M=4,N=2$, C: $M=3,N=3$ and D: $M=4, N=3$.}
\end{figure*}

\begin{figure*}[ht]
\centering
\includegraphics*[width=0.92 \textwidth]{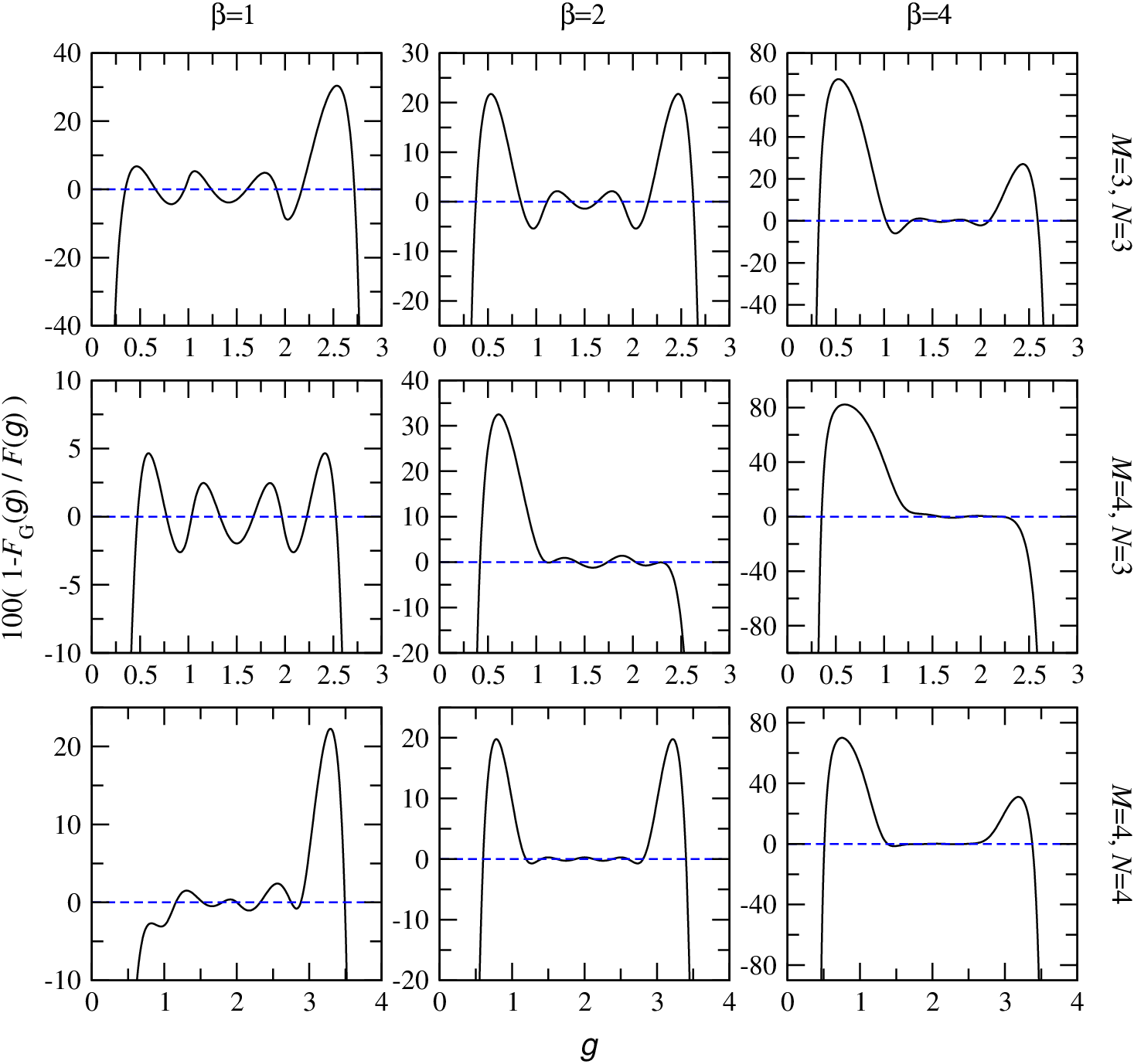}
\caption[]{Relative percentage difference between exact distribution and Gaussian approximation.}
\end{figure*}

\section {Conductance Distribution for Crossover Ensembles}
\label{secCDCE}

\subsection {Crossover Ensembles}

The OE-UE and SE-UE transitions are important for studying the effect of magnetic field on the mesoscopic cavities \cite{CBP,CCM,SKP,PW,FP}. Variation of the magnetic field leads to crossover from the OE or SE (time-reversal invariant) to UE (time-reversal noninvariant). This in turn gives rise to the phenomenon of weak-localization or weak-antilocalization in crossover ensembles. We have recently worked out the statistics of transmission eigenvalues for both OE-UE and SE-UE crossover ensembles for arbitrary $N_1,N_2$ \cite{SKP}. The crossover is governed by a symmetry breaking parameter $\tau$. The transitions from OE to UE and SE to UE take place as $\tau$ is varied from 0 to $\infty$. 

The conductance problem in chaotic cavities has also been analysed using the Hamiltonian formalism as a microscopic justification for the scattering matrix approach \cite{Bnkr,PW}. To consider the crossover regime under this approach, one assumes the Hamiltonian to belong to the Gaussian crossover ensemble of random matrices \cite{FP}. For instance, for the OE-UE crossover the Hamiltonian is taken as \cite{PW,FP,PM}
\begin{equation}
 H_\eta=H_0(v^2)+i \eta A(v^2).
\end{equation}
 Here $H_0$ belongs to GOE and has $v^2$ as the variance for non-diagonal elements. $A$ is an antisymmetric matrix whose independent matrix elements are Gaussians with variance $v^2$. $\eta$ serves as the transition parameter with $\eta=0$ corresponding to GOE and $\eta=1$ corresponding to GUE. Using various theoretical arguments, supported by numerical evidences, it has been shown that the parameter $\eta$ is proportional to the magnetic flux through the cavity \cite{BGS,PW}. The connection of the parameter $\tau$ with the magnetic flux has been established via the parameter $\eta$ by relating the scattering matrix to the Hamiltonian \cite{Bnkr,FP}. It turns out that, for the range of magnetic field strength for which theory holds good, the parameter $\tau$ is proportional to the square of magnetic flux through the system. A comparison of random matrix results \cite{SKP,FP} for averages of conductance and shot-noise power with the corresponding semiclassical results \cite{HMBH,HMH,BC} also leads to the same conclusion. 
 
In \cite{SKP} we have obtained the averages and variances of conductance and shot-noise power for the crossover ensembles. We show here that the Laplace transform method can be used to derive the conductance distributions in the crossover ensembles also.

The JPD of transmission eigenvalues for the crossover is given by \cite{SKP}
\begin{equation}
\label{JPDt}
P(T_1,...T_N;\tau)=e^{\mathcal{E}_0\tau}C_N\Delta_N(T_1,...,T_N)\mbox{Pf}[\mathcal{F}_{j,k}]\prod_{l=1}^N T_l^{\gamma}, 
\end{equation}
where $C_N=C_N^{(1)}$ for OE-UE crossover and $C_N^{(4)}$ for SE-UE crossover. $\mathcal{E}_0$ is given by
\begin{equation}
\mathcal{E}_0=\sum_{\mu=0}^{N-1} \varepsilon_\mu ,~~~~~~~~~\varepsilon_\mu=\mu(\mu+2b+2).
\end{equation}
The parameter $b$ is defined by
\begin{equation}
2b+1=|N_1-N_2|.
\end{equation}
Also $\gamma=b$ and $b+1$ respectively for the OE-UE and SE-UE crossovers. The crossover parameter $\tau$ appears in (\ref{JPDt}) in the normalization and the Pfaffian. $\mathcal{F}_{j,k}$ in (\ref{JPDt}) is an antisymmetric function with the indices $j,k$  taking values from 1 to $N$ or $N+1$ depending on whether $N$ is even or odd. $N$ is necessarily even in the SE-UE crossover to take care of Kramers degeneracy explicitly for SE. For $j,k=1,2,...,N$, $\mathcal{F}_{j,k}=G(T_j,T_k;\tau)$. In addition, for odd $N$ case of OE-UE crossover $\mathcal{F}_{j,N+1}=-\mathcal{F}_{N+1,j}=H(T_j;\tau)(1-\delta_{j,N+1})$. The explicit forms of $G$ and $H$ have been given below. We use the above JPD and other results from \cite{SKP} to obtain the conductance distributions for the crossover ensembles. 

As in (\ref{LF-beta}) the Laplace transform of the conductance distribution is given by
\begin{equation}
\widetilde{F}(s;\tau)=\int_0^1\cdots\int_0^1 P(T_1,...T_N;\tau)\prod_{j=1}^N e^{-sT_j}\,dT_1...dT_N.
\end{equation}
The above integral can be evaluated by expanding the Pfaffian and then performing the integral using the method of alternate variables \cite{RM,PM,MP}. The final answer is obtained in terms of a Pfaffian. We outline the proof in appendix A. We give the results for the OE-UE and SE-UE crossovers below. 
\subsection {OE-UE Crossover}

For the OE-UE crossover we get the Laplace transform of the conductance distribution as
\begin{equation}
\label{LF-OUe}
\widetilde{F}(s;\tau)=e^{\mathcal{E}_0\tau}\Gamma(N+1)C_N^{(1)}\mbox{Pf}[\Psi_{j,k}(s;\tau)]_{j,k=0,...,N-1},
\end{equation}
when $N$ is even, and
\begin{equation}
\label{LF-OUo}
\widetilde{F}(s;\tau)=e^{\mathcal{E}_0\tau}\Gamma(N+1)C_N^{(1)}\mbox{Pf}
            \left[\begin{array}{cc}
             \Psi_{j,k}(s;\tau) & \Phi_j(s;\tau) \\
            -\Phi_k(s;\tau)  &  0
            \end{array}\right]_{j,k=0,...,N-1},        
\end{equation} 
when $N$ is odd. In (\ref{LF-OUe}) and (\ref{LF-OUo}) $\Psi$ and $\Phi$ are given by
\begin{equation}
\label{Psi-OU}
\Psi_{j,k}(s;\tau)=\int_{0}^1\int_{0}^1 G(x,y;\tau)\,e^{-sx}e^{-sy}x^{\gamma+j}y^{\gamma+k}\,dx\,dy,
\end{equation}
\begin{equation}
\label{Phi-OU}
\Phi_j(s;\tau)=\int_0^1 H(x;\tau)e^{-sx}x^{\gamma+j} dx,
\end{equation}
where 
\begin{eqnarray}
\fl
\nonumber 
G(x,y;\tau)=4x^{b+1}y^{b+1}\sum_{\mu=0}^\infty\sum_{\nu=\mu+1}^\infty(-1)^{\mu+\nu}e^{-(\varepsilon_\mu+\varepsilon_\nu)\tau}\\
  \nonumber
 \times\left[P_\mu^{0,2b+1}(2x-1)P_\nu^{0,2b+1}(2y-1)-P_\nu^{0,2b+1}(2x-1)P_\mu^{0,2b+1}(2y-1)\right]\\
 \nonumber
 =-4x^{b+1}y^{b+1}\sum_{\mu=1}^\infty\sum_{\nu=0}^{\mu-1}(-1)^{\mu+\nu}e^{-(\varepsilon_\mu+\varepsilon_\nu)\tau}\\
 \times\left[P_\mu^{0,2b+1}(2x-1)P_\nu^{0,2b+1}(2y-1)-P_\nu^{0,2b+1}(2x-1)P_\mu^{0,2b+1}(2y-1)\right],
\end{eqnarray}
\begin{eqnarray}
\fl
H(x;\tau)=2x^{b+1}\sum_{\mu=0}^{\infty}(-1)^\mu e^{-\varepsilon_\mu\tau}P_\mu^{0,2b+1}(2x-1). 
\end{eqnarray}
Here the $P_j^{a,b}(x)$ are Jacobi polynomials. Note that in the $\tau\rightarrow 0$ limit $G(x,y;\tau)=\mbox{sgn}(x-y)$ and $H(x;\tau)=1$. The OE result therefore follows from the above crossover result for $\tau=0$.

\begin{figure*}[ht]
\centering
\includegraphics*[width=0.92 \textwidth]{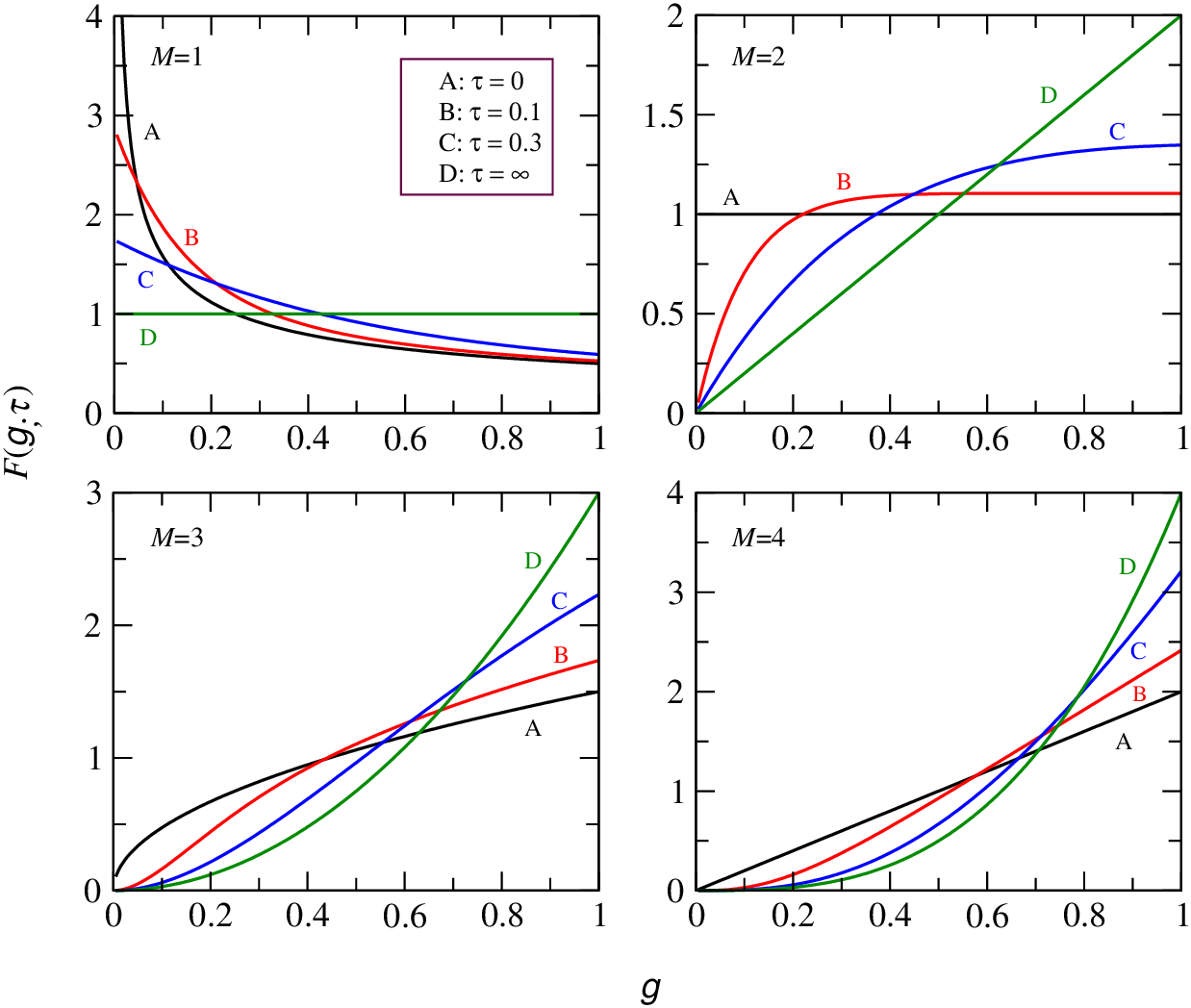}
\caption[]{Conductance distribution in OE-UE crossover for $N=1$ and $M=1,2,3,4$, with different values of $\tau$.}
\end{figure*}

For arbitrary $N_1,N_2$ the results for conductance distribution are complicated. However, for $N=1$ and arbitrary $M$ the final expression is simple. We find
\begin{equation}
F(g;\tau)=Mg^{M-1}\sum_{\nu=0}^\infty (-1)^\nu e^{-\nu(\nu+M)\tau} P_\nu^{0,M-1}(2g-1).
\end{equation}
 It is easy to see that $\tau\rightarrow\infty$ reproduces the correct UE result. For the other limit $\tau=0$, which gives the OE result, one has to use the following identity which follows from the generating function for Jacobi polynomials \cite{Sz}:
\begin{equation}
\sum_{\nu=0}^\infty (-1)^\nu P_\nu^{0,M-1}(2g-1)=\frac{1}{2}g^{-M/2}.
\end{equation}
The effect of varying $\tau$ on conductance distribution has been shown in figure 4 for $M\leq4$.

\subsection {SE-UE Crossover}
For the SE-UE crossover the Laplace transform of the conductance distribution is given by
\begin{equation}
\label{LF-SU}
\widetilde{F}(s;\tau)=e^{\mathcal{E}_0\tau}\Gamma(N+1)C_N^{(4)}\mbox{Pf}[\Psi_{j,k}(s;\tau)]_{j,k=0,...,N-1},
\end{equation}
where as in (\ref{LF-OUe}),
\begin{equation}
\label{Psi-SU}
\Psi_{j,k}(s;\tau)=\int_{0}^1\int_{0}^1 G(x,y;\tau)\,e^{-sx}e^{-sy}x^{\gamma+j}y^{\gamma+k}\,dx\,dy.
\end{equation}
In this case $G(x,y;\tau)$ has the expansion
\begin{eqnarray}
\fl
\nonumber G(x,y;\tau)=-\frac{1}{2}x^{b}y^{b}\sum_{\mu=0}^\infty\sum_{\nu=0}^\infty(\mu+b+1)(\nu+b+1)e^{-(\varepsilon_\mu+\varepsilon_\nu)\tau}\\
  \nonumber
 \times\left[P_\mu^{0,2b+1}(2x-1)P_\nu^{0,2b+1}(2y-1)-P_\nu^{0,2b+1}(2x-1)P_\mu^{0,2b+1}(2y-1)\right]\\
 \nonumber
 =\frac{1}{2}x^{b}y^{b}\sum_{\mu=1}^\infty\sum_{\nu=0}^{\mu-1}(\mu+b+1)(\nu+b+1)e^{-(\varepsilon_\mu+\varepsilon_\nu)\tau}\\
 \times\left[P_\mu^{0,2b+1}(2x-1)P_\nu^{0,2b+1}(2y-1)-P_\nu^{0,2b+1}(2x-1)P_\mu^{0,2b+1}(2y-1)\right].
\end{eqnarray}
For $\tau\rightarrow 0$ we have $G(x,y;\tau)=-\partial\delta(x-y)/\partial x$. The SE result given in section 3 does not take into account Kramers degeneracy. One therefore has to properly scale the quantities to obtain (\ref{LF4}) from (\ref{LF-SU}) in the $\tau=0$ limit.

\subsection{Large \texorpdfstring{$N_1, N_2$}{N1, N2} results}

For large $N$ the conductance distribution is expected to behave like a Gaussian as in (\ref{Gauss}). However, the average and variance should be appropriate to the crossover ensembles, as given in \cite{SKP}. We have for arbitrary $N_1,N_2$,
\begin{eqnarray}
\label{Gt-O}
\left<g\right>=\frac{N_1 N_2}{N_\mathrm{s}}-\frac{N_1 N_2}{N_\mathrm{s}(N_\mathrm{s}+1)}e^{-N_\mathrm{s}\tau},
\end{eqnarray}
and
\begin{eqnarray}
\label{varGt-O}
\fl
\nonumber
\mbox{var}(g)=\frac{N_1^2N_2^2}{(N_\mathrm{S}-1)(N_\mathrm{S})^2(N_\mathrm{S}+1)}+\frac{N_1 N_2(N_1-1)(N_2-1)}{(N_\mathrm{S}-2)(N_\mathrm{S}-1)(N_\mathrm{S})(N_\mathrm{S}+1)}e^{-2(N_\mathrm{S}-1)\tau}\\
\nonumber
-\frac{N_1^2 N_2^2}{(N_\mathrm{S})^2(N_\mathrm{S}+1)^2}e^{-2N_\mathrm{S}\tau}+\frac{N_1 N_2(N_1+1)(N_2+1)}{N_\mathrm{S}(N_\mathrm{S}+1)(N_\mathrm{S}+2)(N_\mathrm{S}+3)}e^{-2(N_\mathrm{S}+1)\tau}\\
+\frac{2N_1 N_2(N_1-N_2)^2}{(N_\mathrm{S}-2)(N_\mathrm{S})^2(N_\mathrm{S}+1)(N_\mathrm{S}+2)}e^{-N_\mathrm{S}\tau},
\end{eqnarray}
for OE-UE crossover. Similarly
\begin{eqnarray}
\label{Gt-S}
\left<g\right>=\frac{N_1 N_2}{N_\mathrm{S}}+\frac{N_1 N_2}{N_\mathrm{S}(N_\mathrm{S}-1)}e^{-N_\mathrm{S}\tau},
\end{eqnarray}
and
\begin{eqnarray}
\label{varGt-S}
\fl
\mbox{var}(g)=\frac{N_1^2N_2^2}{(N_\mathrm{S}-1)(N_\mathrm{S})^2(N_\mathrm{S}+1)}+\frac{N_1 N_2(N_1-1)(N_2-1)}{(N_\mathrm{S}-3)(N_\mathrm{S}-2)(N_\mathrm{S}-1)(N_\mathrm{S})}e^{-2(N_\mathrm{S}-1)\tau}\nonumber\\
-\frac{N_1^2 N_2^2}{(N_\mathrm{S}-1)^2(N_S)^2}e^{-2N_\mathrm{S}\tau}+\frac{N_1 N_2(N_1+1)(N_2+1)}{(N_\mathrm{S}-1)(N_\mathrm{S})(N_\mathrm{S}+1)(N_\mathrm{S}+2)}e^{-2(N_\mathrm{S}+1)\tau}\nonumber\\
-\frac{2N_1 N_2(N_1-N_2)^2}{(N_\mathrm{S}-2)(N_\mathrm{S}-1)(N_\mathrm{S})^2(N_\mathrm{S}+2)}e^{-N_\mathrm{S}\tau},
\end{eqnarray}
for SE-UE crossover. For large $N_1, N_2$ the above equations simplify to
\begin{equation}
\label{avG_t}
\left<g\right>=\frac{N_1 N_2}{N_\mathrm{S}}\left(1\mp\frac{e^{-N_\mathrm{S}\tau}}{N_\mathrm{S}}\right),
\end{equation}
and 
\begin{equation}
\label{varG_t}
\mbox{var}(g)=\frac{N_1^2 N_2^2}{N_\mathrm{S}^4}\left(1+e^{-2N_\mathrm{S}\tau}\right),
\end{equation}
valid for both the crossovers. In (\ref{avG_t}) the upper and lower signs correspond respectively to the OE-UE and SE-UE crossovers. It is clear from (\ref{avG_t}) and (\ref{varG_t}) that for large $N_1,N_2$ while the average changes very little for both the crossovers, the variance becomes half as $\tau$ varies from 0 to $\infty$.

\section {Conclusion}
\label{secC}

To conclude we have proposed a formalism to obtain exact distributions of conductance in chaotic mesoscopic cavities for all the three invariant classes of random matrices. The technique is particularly useful for finding explicit answers for small $N$ values where one expects significant deviation from Gaussian-like behavior. These results are important from the point of view of experiments where similar number of channels are typically considered.  

We have also worked out the conductance distributions for the OE-UE and SE-UE crossovers using the results of \cite{SKP}. These results are important for investigating the behaviour of conductance distributions in mesoscopic cavities with small magnetic field. 

We remark that, working in similar fashion, it is possible to present the exact distribution of shot-noise power also as an inverse Laplace transform of determinant or Pfaffian. One just needs to replace $e^{-sx}$ by $e^{-sx(1-x)}$ in the expressions for $\Psi$ and $\Phi$. However, evaluation of the inverse Laplace transform to obtain explicit results poses technical difficulties.

\ack
S. K. acknowledges CSIR India for financial assistance.

\appendix
\section{Proofs of \texorpdfstring{(\ref{LF-OUe})}{(52)}, \texorpdfstring{(\ref{LF-OUo})}{(53)} and \texorpdfstring{(\ref{LF-SU})}{(60)}}
\setcounter{section}{1}

The Pfaffian of a $2\mu \times 2\mu$ antisymmetric matrix $\mathcal{A}$ is defined as \cite{RM}
\begin{equation}
\label{Pf}
\mbox{Pf}[\mathcal{A}]=\sum_{\mathrm{p}} \sigma_{\mathrm{p}} \mathcal{A}_{i_1,i_2}\mathcal{A}_{i_3,i_4}\cdots \mathcal{A}_{i_{2\mu-1},i_{2\mu}}.
\end{equation}
The sum in (\ref{Pf}) is over all permutations $$\mbox{p}={1,~ 2,... ,~ 2\mu \choose i_1,~i_2,...,~i_{2\mu}}$$ with the restrictions $i_1<i_2,~i_3<i_4,~...,~i_{2\mu-1}<i_{2\mu}$; $i_1<i_3<...<i_{2\mu-1}$ and $\sigma_{\mathrm{p}}$ is sign of the permutation. Also, Pfaffian is related to the determinant as 
\begin{equation}
\det[\mathcal{A}]=(\mbox{Pf}[\mathcal{A}])^2.
\end{equation}

For the OE-UE crossover, Laplace transform of conductance distribution is
\begin{eqnarray}
\fl
\nonumber
 \widetilde{F}(s)=e^{\mathcal{E}_0\tau}C_N^{(1)}\int_0^1\,dT_1\cdots\int_0^1\,dT_N \Big(\prod_{i=1}^N e^{-sTi}T_i^\gamma\Big)\mbox{det}[T_k^j]_{^{j=0,..,N-1}_{ k=1,..,N}}\mbox{Pf}[\mathcal{F}_{m,n}]\\
 =e^{\mathcal{E}_0\tau}C_N^{(1)}\int_0^1\,dT_1\cdots\int_0^1\,dT_N\,\mbox{det}[e^{-sT_j}T_k^{j+\gamma}]_{^{j=0,..,N-1}_{ k=1,..,N}}\mbox{Pf}[\mathcal{F}_{m,n}],
\end{eqnarray}
where $m,n=1,...,N$ or $N+1$ depending on whether $N$ is even or odd.

When $N$ is even, the expansion for Pfaffian and symmetry of the $T$ variables lead to
\begin{eqnarray}
\fl
\nonumber
 \widetilde{F}(s)=(N-1)!!\,e^{\mathcal{E}_0\tau}C_N^{(1)}\int_0^1\!\!dT_1\cdots\int_0^1\!\!dT_N \Big(\prod_{i=1}^N e^{-sTi}T_i^\gamma\Big)\mbox{det}[T_k^j]_{^{j=0,..,N-1}_{ k=1,..,N}}(\mathcal{F}_{1,2}\mathcal{F}_{3,4}...\mathcal{F}_{N-1,N})
 \end{eqnarray}
 \begin{eqnarray}
 \fl
 \nonumber
 =(N-1)!!\,e^{\mathcal{E}_0\tau}C_N^{(1)}\int_0^1\!\! dT_1\cdots\int_0^1\!\!dT_N \mbox{det}[e^{-sT_{2k-1}}T_{2k-1}^{\gamma+j}\mathcal{F}_{2k-1,2k} ~~~e^{-sT_{2k}}T_{2k}^{\gamma+j}]_{^{j=0,..,N-1}_{k=1,..,N/2}}~.
 \end{eqnarray}
Defining
\begin{equation}
f_{2k}^\nu=\int_0^1 \,dT_{2k-1} e^{-sT_{2k-1}}T_{2k-1}^{\nu}\mathcal{F}_{2k-1,2k},
\end{equation}
and integrating over the odd numbered columns in the determinant we get
\begin{eqnarray}
\label{altv1}
 \fl
 \widetilde{F}(s)=(N-1)!!\,e^{\mathcal{E}_0\tau}C_N^{(1)}\int_0^1\!\!dT_2\int_0^1\!\!dT_4\cdots\int_0^1\!\!dT_N \mbox{det}[f_{2k}^{\gamma+j} ~~~e^{-sT_{2k}}T_{2k}^{\gamma+j}]_{^{j=0,..,N-1}_{k=1,..,N/2}}.
\end{eqnarray}
The remaining integrals give rise to a Pfaffian \cite{RM,PM,MP}, viz.,
\begin{eqnarray}
 \fl
\nonumber
 \widetilde{F}(s)=(N-1)!!\,e^{\mathcal{E}_0\tau}C_N^{(1)} 2^{N/2}(N/2)! \mbox{Pf}[\Psi_{j,k}(s;\tau)]_{j,k=0,...,N-1}\\
 =e^{\mathcal{E}_0\tau}\Gamma(N+1)C_N^{(1)}\mbox{Pf}[\Psi_{j,k}(s;\tau)]_{j,k=0,...,N-1}, 
\end{eqnarray}
where $\Psi_{j,k}(s;\tau)$ is given by (\ref{Psi-OU}).

When $N$ is odd we get instead of (\ref{altv1}),
\begin{eqnarray}
\label{altv2}
 \fl
 \widetilde{F}(s)=N!!\,e^{\mathcal{E}_0\tau}C_N^{(1)}\int_0^1\!\!\!dT_2\!\int_0^1\!\!\!dT_4\cdots\!\!\int_0^1\!\!\!dT_{N+1} \mbox{det}[f_{2k}^{\gamma+j} ~~~e^{-sT_{2k}}T_{2k}^{\gamma+j},f_{N+1}^{\gamma+j} ]_{^{j=0,..,N-1}_{k=1,..,(N-1)/2}}.
\end{eqnarray}
This then leads to (\ref{LF-OUo}).

The proof for SE-UE crossover result is similar to even $N$ case of OE-UE crossover. Note that the proofs for the $\beta=1,4$ invariant ensembles are implicit in the above derivations.


\end{document}